\documentstyle[12pt,a4]{article}

\begin{document}

\begin{titlepage}
\begin{flushright}
{\large \bf UCL-IPT-97-11}
\end{flushright}
\vskip 2cm
\begin{center}
\vskip .2in

{\Large \bf
Equivalence of the Sine-Gordon and Massive Thirring 
Models at Finite Temperature}
\vskip .4in

{\large D. Del\'epine,  R. Gonz\'alez Felipe and  J. Weyers}\\[.15in]

{\em Institut de Physique Th\'eorique}\\

{\em Universit\'e catholique de Louvain}\\

{\em  B-1348 Louvain-la-Neuve, Belgium}

\end{center}

\vskip 1.5in

\begin{abstract}
Using the path-integral approach, the quantum massive Thirring and sine-Gordon
models are proven to be equivalent at finite temperature. This result is an extension 
of Coleman's proof of the equivalence between both theories at zero temperature.
The usual identifications among the parameters of these models also remain valid
 at $T \neq 0$.

\end{abstract}
\end{titlepage}

\section{Introduction}

In two dimensional quantum field theories fermionic degrees of freedom can
be expressed as bosonic ones and vice versa. A remarkable illustration of
this property is the equivalence between the sine-Gordon and (massive)
Thirring models\cite{coleman}. The sine-Gordon model is a (1+1)-dimensional
field theory of a single scalar field, whose Lagrangian density in Euclidean
space-time is defined classically by 
\begin{equation}
{\cal L}_{SG}=\frac{1}{2}\partial _{\mu }\varphi \partial _{\mu }\varphi -%
\frac{\alpha _{0}}{\lambda ^{2}}\left( \cos \lambda \varphi -1\right) ,
\label{1}
\end{equation}
where $\alpha _{0}$ plays the role of a squared mass, $\lambda $ is a
coupling constant and the minimum energy is taken to be zero. On the other
hand, the Lagrangian density of the massive Thirring model in Euclidean
(1+1)-dimensional space-time is given by 
\begin{equation}
{\cal L}_{T}=-i\bar{\psi}\not{\partial}\psi -\frac{1}{2}g^{2}(\bar{\psi}%
\gamma ^{\mu }\psi )^{2}+imz\bar{\psi}\psi ,  \label{2}
\end{equation}
where $z$ is a cutoff-dependent constant and the $\gamma _{\mu }$ matrices
are taken in the form: 
\begin{equation}
\gamma _{0}=\left( 
\begin{array}{cc}
0 & 1 \\ 
1 & 0
\end{array}
\right) ,\gamma _{1}=\left( 
\begin{array}{cc}
0 & i \\ 
-i & 0
\end{array}
\right) ,\gamma _{5}=i\gamma _{0}\gamma _{1}.  \label{3}
\end{equation}

The equivalence between the two models was first derived by Coleman\cite
{coleman} following Klaiber's work on the massless Thirring model\cite
{klaiber}: he showed that the perturbation series in the mass parameter $m$
of the Thirring model is term-by-term identical with a perturbation series
in $\alpha _{0}$ for the sine-Gordon model, provided the following
identifications are made\cite{coleman,mandelstam}: 
\begin{equation}
\frac{4\pi }{\lambda ^{2}}=1+\frac{g^{2}}{\pi }\ ,  \label{4}
\end{equation}
\begin{equation}
\bar{\psi}\gamma _{\mu }\psi =i\frac{\lambda }{2\pi }\epsilon _{\mu \nu
}\partial _{\nu }\varphi ,  \label{5}
\end{equation}
\begin{equation}
imz\bar{\psi}\psi =-\frac{\alpha _{0}}{\lambda ^{2}}\cos (\lambda \varphi ).
\label{6}
\end{equation}
It is worth noticing that relation (\ref{4}) expresses a {\em duality
symmetry} between the Thirring and sine-Gordon models, i.e. a correspondence
that relates the strong coupling regime in one theory to the weak coupling
one in the other.

Different approaches have been used to prove the equivalence between the
above models. Mandelstam\cite{mandelstam} rederived Coleman's results by
explicitly constructing the operators for creation and annihilation of
quantum sine-Gordon solitons. These operators satisfy the anticommutation
relations and field equations of the fermionic fields in the massive
Thirring model. Alternatively, in the path-integral framework, Coleman's
proof can be rederived in a very simple\cite{naon} way by making a chiral
transformation in the fermionic path-integral variables.

The results quoted above have been obtained in the usual context of
relativistic quantum field theory, namely at zero temperature. An obvious
question is whether they remain valid at finite temperature $T$. Although a
positive answer might be expected from general arguments (bosonisation at
finite $T$) contradictory results have been presented in the literature\cite
{wolf}-\cite{fujita}.

In this letter we give an explicit proof in the path integral approach\cite
{naon} of the equivalence between the massive Thirring and sine-Gordon
models at finite temperature in precisely the terms of Coleman's original
argument at zero temperature. Namely, we show that a perturbative expansion
in the mass of the Thirring model is term-by-term identical with a
perturbation series in $\alpha _{0}$ of the sine-Gordon model, provided the
identifications given in Eqs.(\ref{4})-(\ref{6}) are made.

\section{The massive Thirring and sine-Gordon models at finite temperature}

In this section we shall evaluate the partition functions of the massive
Thirring and sine-Gordon models using the imaginary time formalism and the
path integral approach at finite temperature\cite{bernard}.

Let us first consider the massive Thirring model. The Euclidean partition
function reads 
\begin{equation}
Z_{T}=N_{0}N_{\beta }\int {\cal D}\psi {\cal D}\bar{\psi}e^{-\int d^{2}x%
{\cal L}_{T}},  \label{7}
\end{equation}
where $\int d^{2}x\equiv \int_{0}^{\beta }dx_{0}\int_{-\infty }^{+\infty
}dx_{1}$($\beta =1/T$) and ${\cal L}_{T}$ is given by Eq.(\ref{2}). $N_{0} $
is an infinite temperature-independent normalization constant, whereas $%
N_{\beta }$ is a divergent temperature-dependent constant to be determined
from the free partition function\cite{bernard}. The functional integral is
performed over fermionic fields satisfying antiperiodic boundary conditions
in the time direction,

\begin{equation}
\psi (x_{0},x_{1})=-\psi (x_{0}+\beta ,x_{1}).  \label{8}
\end{equation}

The first step of the evaluation is to introduce an auxiliary two-component
vector field\cite{naon} 
\begin{equation}
A_{\mu }=-\frac{1}{g}(\epsilon _{\mu \nu }\partial _{\nu }\phi -\partial
_{\mu }\eta ),  \label{9}
\end{equation}
where $\phi $ and $\eta $ are two scalar fields satisfying the periodic
boundary conditions

\begin{equation}
\phi (x_{0},x_{1})=\phi (x_{0}+\beta ,x_{1})\ ,\ \eta (x_{0},x_{1})=\eta
(x_{0}+\beta ,x_{1}).  \label{10}
\end{equation}
With the help of these fields, the quartic interaction in ${\cal L}_{T}$ can
be eliminated,

\[
\exp \left( \frac{g^{2}}{2}\int d^{2}x(\bar{\psi}\gamma _{\mu }\psi
)^{2}\right) =\int {\cal D}A_{\mu }\exp \left( -\int d^{2}x(\frac{1}{2}%
A_{\mu }^{2}-g\bar{\psi}\not{A}\psi )\right) , 
\]
and we obtain the effective Lagrangian 
\begin{equation}
{\cal L}_{T}=-\bar{\psi}[i\not{\partial}-\gamma _{\mu }(\epsilon _{\mu \nu
}\partial _{\nu }\phi -\partial _{\mu }\eta )]\psi +imz\bar{\psi}\psi +\frac{%
1}{2g^{2}}[(\partial _{\mu }\phi )^{2}+(\partial _{\mu }\eta )^{2}].
\label{11}
\end{equation}

We now perform the chiral transformation

\begin{eqnarray}
\psi (x) &=&e^{\gamma _{5}\phi (x)+i\eta (x)}\chi (x),  \nonumber \\
\bar{\psi}(x) &=&\bar{\chi}(x)e^{\gamma _{5}\phi (x)-i\eta (x)},  \label{12}
\end{eqnarray}
with $\chi (x)$ satisfying the boundary condition

\begin{equation}
\chi (x_{0},x_{1})=-\chi (x_{0}+\beta ,x_{1}),  \label{13}
\end{equation}
to obtain 
\begin{equation}
{\cal L}_{T}=-\bar{\chi}i\not{\partial}\chi +imz\bar{\chi}e^{2\gamma
_{5}\phi (x)}\chi +\frac{1}{2g^{2}}[(\partial _{\mu }\phi )^{2}+(\partial
_{\mu }\eta )^{2}].  \label{14}
\end{equation}

To write the partition function in terms of the new variables, the Jacobians
of the transformations 
\begin{eqnarray}
{\cal D}\psi {\cal D}\bar{\psi} &=&J_{F}{\cal D}\chi {\cal D}\bar{\chi}, 
\nonumber \\
{\cal D}A_{\mu } &=&J_{A}{\cal D}\phi {\cal D}\eta .  \label{15}
\end{eqnarray}
have to be properly taken into account. Due to the anomaly and the fact that
we perform a chiral transformation, the first Jacobian $J_{F}$ is not
trivial. It has been computed at finite temperature in Ref.\cite{reuter}
following Fujikawa's procedure\cite{fujikawa} and the result is 
\begin{equation}
J_{F}=e^{-\frac{1}{2\pi }\int d^{2}x(\partial _{\mu }\phi )^{2}}.  \label{16}
\end{equation}
The bosonic Jacobian $J_{A}$ on the other hand is easily evaluated and one
finds

\begin{equation}
J_{A}=\det \frac{-\nabla ^{2}}{g^{2}},  \label{17}
\end{equation}
where $\nabla ^{2}=\partial _{\mu }\partial _{\mu }.$ Note that this bosonic
determinant is temperature-dependent and hence its contribution is relevant
to the partition function, in contrast to the zero-temperature case where it
plays no role and can simply be absorbed in the normalization constant.

Finally we have,

\begin{equation}
Z_{T}=N_{0}N_{\beta }J_{A}\int {\cal D}\chi {\cal D}\bar{\chi}{\cal D}\phi 
{\cal D}\eta e^{-\int d^{2}x{\cal L}_{T}},  \label{18}
\end{equation}
with 
\begin{equation}
{\cal L}_{T}=-\bar{\chi}i\not{\partial}\chi +imz\bar{\chi}e^{2\gamma
_{5}\phi (x)}\chi +\frac{1}{2\kappa ^{2}}(\partial _{\mu }\phi )^{2}+\frac{1%
}{2g^{2}}(\partial _{\mu }\eta )^{2}  \label{19}
\end{equation}
and

\begin{equation}
\kappa ^{2}=\frac{g^{2}}{1+g^{2}/\pi }\ .  \label{20}
\end{equation}
The integration is performed over fields satisfying the boundary conditions (%
\ref{10}) and (\ref{13}). As can be seen, the $\eta $ field decouples
completely and thus can be trivially integrated out.

In order to show the equivalence between the massive Thirring and the
sine-Gordon models, let us expand $Z_{T}$ in the mass parameter $zm$, 
\begin{eqnarray}
Z_{T} &=&N_{0}N_{\beta }J_{A}^{1/2}\int {\cal D}\chi {\cal D}\bar{\chi}{\cal %
D}\phi e^{-\int d^{2}x(-\bar{\chi}i\not{\partial}\chi +\frac{1}{2\kappa ^{2}}%
(\partial _{\mu }\phi )^{2})}  \nonumber \\
&&\sum_{n=0}^{\infty }\frac{(-izm)^{n}}{n!}\prod_{j=1}^{n}\int d^{2}x_{j}%
\bar{\chi}(x_{j})e^{2\gamma _{5}\phi (x_{j})}\chi (x_{j}).  \label{21}
\end{eqnarray}
Or, equivalently,

\begin{equation}
Z_{T}=Z_{FD}\det \left( \frac{\kappa }{g}\right) \sum_{n=0}^{\infty }\frac{%
(-izm)^{n}}{n!}\left\langle \prod_{j=1}^{n}\int d^{2}x_{j}\bar{\chi}%
(x_{j})e^{2\gamma _{5}\phi (x_{j})}\chi (x_{j})\right\rangle ,  \label{22}
\end{equation}
where $\left\langle \;\right\rangle $ denotes the thermal average over the
unperturbed ensemble and $Z_{FD}$ is the Fermi-Dirac distribution for
massless fermions\cite{bernard}:

\begin{equation}
\ln Z_{FD}=2\int\limits_{0}^{\infty }\frac{dk}{2\pi }\left\{ \frac{\beta k}{2%
}+\ln \left( 1+e^{-\beta k}\right) \right\} .  \label{23}
\end{equation}
At this stage, it is worth noticing that Eq.(\ref{22}) reproduces in the
limit $m=0$ the partition function for the massless Thirring model at finite
temperature\cite{ruiz,manias}.

To evaluate the thermal averages in Eq.(\ref{22}) we need the boson and
fermion free propagators at finite temperature. In the imaginary time
formalism and using the Schwinger representation\cite{schwinger}, the
propagator for a scalar with mass $\mu $ can be defined as 
\begin{equation}
D(x-y)=\frac{1}{\beta }\sum_{n=-\infty }^{+\infty }\int \frac{dk_{1}}{2\pi }%
e^{-ik(x-y)}\int_{0}^{\infty }dse^{-s(k^{2}+\mu ^{2})},  \label{24}
\end{equation}
with $k^{2}=k_{0}^{2}+k_{1}^{2}$ and $k_{0}=2\pi n/\beta $ - the Matsubara
frequencies. Eq.(\ref{24}) can be expressed in terms of the modified Bessel
function $K_{0}(z)$, 
\begin{equation}
D(x-y)=\frac{1}{2\pi }\sum_{n=-\infty }^{+\infty }K_{0}(\mu \sqrt{%
(x_{0}-y_{0}-n\beta )^{2}+(x_{1}-y_{1})^{2}}).  \label{25}
\end{equation}
When $\mu \rightarrow 0$, we get from the last equation 
\begin{eqnarray}
D(x-y) &=&-\frac{1}{2\pi }\sum_{n=-\infty }^{+\infty }\ln (\mu \sqrt{%
(x_{0}-y_{0}-n\beta )^{2}+(x_{1}-y_{1})^{2}})  \nonumber \\
&=&-\frac{1}{2\pi }\ln \left( \mu \beta \sqrt{\cosh (\frac{2x_{1}\pi }{\beta 
})-\cos (\frac{2x_{0}\pi }{\beta })}\right) ,  \label{26}
\end{eqnarray}
which, as required, is periodic in $x_{0}$. Note that the small mass $\mu $
has been added to avoid infrared divergences.

Similarly, the fermionic propagator can be defined at finite temperature as 
\begin{eqnarray}
S(x-y) &=&\frac{1}{\beta }\sum_{n=-\infty }^{+\infty }\int \frac{dk_{1}}{%
2\pi }e^{-ik(x-y)}(k_{0}\gamma _{0}+k_{1}\gamma _{1})\int_{0}^{\infty
}ds\;e^{-sk^{2}}  \nonumber \\
&\equiv &i\gamma _{0}S_{0}(x-y)+i\gamma _{1}S_{1}(x-y),  \label{27}
\end{eqnarray}
with $k_{0}=(2n+1)\pi /\beta $. The functions $S_{0}(x)$ and $S_{1}(x)$ can
be expressed as: 
\begin{eqnarray}
S_{0}(x) &=&-\frac{1}{\beta }\frac{\cosh (\frac{x_{1}\pi }{\beta })\sin (%
\frac{x_{0}\pi }{\beta })}{\cosh (\frac{2x_{1}\pi }{\beta })-\cos (\frac{%
2x_{0}\pi }{\beta })}\ ,  \label{28} \\
S_{1}(x) &=&-\frac{1}{\beta }\frac{\sinh (\frac{x_{1}\pi }{\beta })\cos (%
\frac{x_{0}\pi }{\beta })}{\cosh (\frac{2x_{1}\pi }{\beta })-\cos (\frac{%
2x_{0}\pi }{\beta })}\ .  \label{29}
\end{eqnarray}
$S(x-y)$ is thus antiperiodic in $x_{0}$ with a period equal to $\beta $, as
it should be.

The scalar and fermion propagators can be rewritten in a more familiar way,
using the following dimensionless ``generalized coordinates'' $Q\equiv
(Q_{0},Q_{1})$, 
\begin{eqnarray}
Q_{0}(x) &=&-\cosh (\frac{x_{1}\pi }{\beta })\sin (\frac{x_{0}\pi }{\beta }),
\nonumber \\
Q_{1}(x) &=&-\sinh (\frac{x_{1}\pi }{\beta })\cos (\frac{x_{0}\pi }{\beta }).
\label{30}
\end{eqnarray}
and 
\begin{equation}
Q^{2}(x)\equiv Q_{0}^{2}(x)+Q_{1}^{2}(x)=\cosh (\frac{2x_{1}\pi }{\beta }%
)-\cos (\frac{2x_{0}\pi }{\beta }).  \label{31}
\end{equation}
In terms of these generalized coordinates, $D(x)$ and $S(x)$ are now given
by 
\begin{equation}
D(x)=-\frac{1}{2\pi }\ln (\mu \beta |Q(x)|),  \label{32}
\end{equation}
\begin{equation}
S(x)=\frac{i}{\beta }\frac{\not{Q}(x)}{Q^{2}(x)}.  \label{33}
\end{equation}

Introducing the composite operators

\begin{equation}
\sigma ^{+}=\bar{\chi}\frac{1+\gamma _{5}}{2}\chi \ ,\sigma ^{-}=\bar{\chi}%
\frac{1-\gamma _{5}}{2}\chi  \label{34}
\end{equation}
and using the relation 
\begin{equation}
\bar{\chi}e^{2\gamma _{5}\phi }\chi =e^{2\phi }\sigma ^{+}+e^{-2\phi }\sigma
^{-},  \label{35}
\end{equation}
Eq.(\ref{22}) now reads

\begin{eqnarray}
Z_{T} &=&Z_{FD}\det \left( \frac{\kappa }{g}\right) \sum_{n=0}^{\infty }%
\frac{(-izm)^{2n}}{n!^{2}}\prod_{j=1}^{n}\int d^{2}x_{j}d^{2}y_{j}  \nonumber
\\
&&\left\langle e^{-2(\phi (x_{j})-\phi (y_{j}))}\sigma ^{+}(x_{j})\sigma
^{-}(y_{j})\right\rangle .  \label{36}
\end{eqnarray}

To compute the bosonic thermal average, we use Wick's theorem and the
well-known identity 
\begin{equation}
{\cal T}(e^{-i\int d^{2}xj(x)\phi (x)})=:e^{^{-i\int d^{2}xj(x)\phi
(x)}}:e^{-\frac{1}{2}\int d^{2}xd^{2}yj(x)\Delta (x-y)j(y)},  \label{37}
\end{equation}
where ${\cal T}$ is the $x_{0}$-ordering chronological product and $:$ $:$
denotes the normal product; $\Delta (x-y)$ is the propagator of the $\phi $
field and $j(x),$ any space-time function. We have then

\begin{equation}
\left\langle \prod_{j=1}^{n}e^{-2(\phi (x_{j})-\phi (y_{j}))}\right\rangle
_{renorm.}=\frac{\prod_{i>j}^{n}(\rho ^{2}\beta ^{2}\left|
Q(x_{i}-x_{j})\right| \left| Q(y_{i}-y_{j})\right| )^{-\frac{2\kappa ^{2}}{%
\pi }}}{\prod_{i,j}^{n}(\rho \beta \left| Q(x_{i}-y_{j})\right| )^{-\frac{%
2\kappa ^{2}}{\pi }}},  \label{38}
\end{equation}
where $\rho $ is a renormalization scale, $Q(x_{i})$ are the generalized
coordinates defined in Eqs.(\ref{30}) and $\kappa $ is given by Eq.(\ref{20}%
). The presence of the latter factor is due to the kinetic term of the
scalar Lagrangian (cf. Eq.(\ref{19})).

The fermionic average is also easily evaluated, if we recall the identity%
\cite{zinn}

\begin{equation}
\left( -1\right) ^{n+1}\det \frac{1}{f(w_{i}-w_{j}^{^{\prime }})}=\frac{%
\prod_{i<j}\ f(w_{i}-w_{j})f(w_{i}^{^{\prime }}-w_{j}^{^{\prime }})}{%
\prod_{i,j}f(w_{i}-w_{j}^{^{\prime }})},  \label{39}
\end{equation}
which holds only for the analytic functions $f(w)=w$ and $f(w)=\sinh (\alpha
w).$ In our case

\begin{equation}
Q_{0}=\frac{i}{2}(\sinh w-\sinh \bar{w}),\ Q_{1}=-\frac{1}{2}(\sinh w+\sinh 
\bar{w}),\ Q^{2}=\sinh w\sinh \bar{w},  \label{40}
\end{equation}
with $w=\left( x_{1}+ix_{0}\right) \pi /\beta $, and thus

\begin{equation}
\left\langle \prod_{j=1}^{n}\sigma ^{+}(x_{j})\sigma
^{-}(y_{j})\right\rangle =\left( -1\right) ^{n}\frac{\prod_{i>j}^{n}(\beta
^{2}\left| Q(x_{i}-x_{j})\right| \left| Q(y_{i}-y_{j})\right| )^{2}}{%
\prod_{i,j}^{n}(\beta \left| Q(x_{i}-y_{j})\right| )^{2}}.  \label{41}
\end{equation}

Substituting Eqs.(\ref{38}) and (\ref{41}) into $Z_{T}$, we obtain finally 
\begin{eqnarray}
Z_{T} &=&Z_{FD}\det \left( \frac{\kappa }{g}\right) \sum_{n=0}^{\infty }%
\frac{(zm)^{2n}}{n!^{2}}\left( \prod_{j=1}^{n}\int
d^{2}x_{j}d^{2}y_{j}\right)  \nonumber \\
&&\frac{\prod_{i>j}^{n}(\rho ^{2}\beta ^{2}\left| Q(x_{i}-x_{j})\right|
\left| Q(y_{i}-y_{j})\right| )^{2-\frac{2\kappa ^{2}}{\pi }}}{%
\prod_{i,j}^{n}(\rho \beta \left| Q(x_{i}-y_{j})\right| )^{2-\frac{2\kappa
^{2}}{\pi }}}\ .  \label{42}
\end{eqnarray}

To compare $Z_{T}$ with the sine-Gordon model, we have to expand in $\alpha
_{0}$ the sine-Gordon partition function 
\begin{equation}
Z_{SG}=N_{0}N_{\beta }^{\prime }\int {\cal D}\varphi \;e^{-\int d^{2}x{\cal L%
}_{SG}},  \label{43}
\end{equation}
where ${\cal L}_{SG}$ is given in Eq.(\ref{1}) and the integration runs over
scalar fields periodic in the time direction:

\begin{equation}
\varphi (x_{0,}x_{1})=\varphi (x_{0}+\beta ,x_{1}).  \label{44}
\end{equation}
This perturbative expansion in $\alpha _{0}$ yields 
\begin{equation}
Z_{SG}=Z_{BE}\sum_{n=0}^{\infty }\frac{1}{n!^{2}}(\frac{\alpha _{0}}{\lambda
^{2}})^{2n}\left\langle \prod_{j=1}^{n}\int e^{i\lambda \varphi
(x_{j})}e^{-i\lambda \varphi (y_{j})}d^{2}x_{j}d^{2}y_{j}\right\rangle ,
\label{45}
\end{equation}
where

\begin{equation}
\ln Z_{BE}=-\int\limits_{0}^{\infty }\frac{dk}{2\pi }\left\{ \frac{\beta k}{2%
}+\ln \left( 1-e^{-\beta k}\right) \right\}   \label{46}
\end{equation}
is the Bose-Einstein distribution for massless bosons. Comparing Eqs.(\ref
{23}) and (\ref{46}) it is straightforward to check that $Z_{BE}=CZ_{FD}$,
where $C$ is an irrelevant (infinite) constant related to the zero-point
energies.

Evaluating the bosonic thermal average with the help of relation (\ref{37}),
we obtain 
\begin{eqnarray}
Z_{SG} &=&Z_{BE}\sum_{n=0}^{\infty }\frac{1}{n!^{2}}(\frac{\zeta \alpha _{0}%
}{\lambda ^{2}})^{2n}\left( \prod_{j=1}^{n}\int d^{2}x_{j}d^{2}y_{j}\right) 
\nonumber \\
&&\frac{\prod_{i>j}^{n}(\beta ^{2}M^{2}\left| Q(x_{i}-x_{j})\right| \left|
Q(y_{i}-y_{j})\right| )^{\frac{\lambda ^{2}}{2\pi }}}{\prod_{i,j}^{n}(M\beta
\left| Q(x_{i}-y_{j})\right| )^{\frac{\lambda ^{2}}{2\pi }}}\ ,  \label{47}
\end{eqnarray}
with $M$ an arbitrary scale and $\zeta $ is an ultraviolet-cutoff-dependent
coefficient.

Comparing $Z_{T}$ and $Z_{SG}$, we see that the two partition functions are
identical provided the relations 
\begin{eqnarray}
\frac{4\pi }{\lambda ^{2}} &=&1+\frac{g^{2}}{\pi }\ ,\bigskip  \label{48} \\
\mathstrut \frac{\zeta \alpha _{0}}{\lambda ^{2}}\ &=&zm,  \label{49} \\
M &=&\rho  \label{50}
\end{eqnarray}
are satisfied. The first relation is independent of the renormalization
scheme, while the last two equations depend on it and hence have only a
convention-dependent meaning. We also recover Coleman's relations (\ref{5})
and (\ref{6}) between the two theories.

In conclusion, we have shown using the path integral method that the
compactification of the time variable into a circle of radius $\beta =1/T$
preserves the equivalence between the sine-Gordon and massive Thirring
models in Coleman's sense: at fixed radius $\beta $ (or at fixed
temperature), the perturbation series in the mass parameter of the Thirring
model is term-by-term identical with a perturbation series in $\alpha _{0}$
for the sine-Gordon model, provided the identifications given in Eqs.(\ref{4}%
)-(\ref{6}) are made.

\end{document}